\documentclass[10pt,a4wide]{article}

\usepackage{geometry}
\usepackage[numbers]{natbib}

\usepackage{graphicx}
\usepackage{color}
\usepackage{url}
\usepackage{xspace}

\newcommand{\comment}[1]{}

\newcommand{\showfig}[1]{#1}
\graphicspath{{./figures/}}

\begin{document}
\title{Logarithmic distributions prove that intrinsic learning is Hebbian%
$^1$
}
\author{Gabriele Scheler\\
Carl Correns Foundation for Mathematical Biology\\
1030 Judson Dr., Mountain View, Ca 94040\\
{\small\tt gscheler@gmail.com}}
\date{}
\maketitle
\footnotetext[1]{
This paper appeared in F1000Research 2017, 6:1222.
The GNN simulation software is available at
\url{https://github.com/gscheler/GNN}.}

\begin{abstract}
In this paper, we present data for the lognormal distributions of spike rates, synaptic weights and intrinsic excitability (gain) for neurons in various brain areas, such as auditory or visual cortex, hippocampus, cerebellum, striatum, midbrain nuclei. We find a remarkable consistency of heavy-tailed, specifically lognormal, distributions for rates, weights and gains in all brain areas examined.
The difference between strongly recurrent and feed-forward connectivity (cortex vs. striatum and cerebellum), neurotransmitter (GABA (striatum) or glutamate (cortex)) or the level of activation (low in cortex, high in Purkinje cells and midbrain nuclei) turns out to be irrelevant for this feature. Logarithmic scale distribution of weights and gains appears to be a general, functional property in all cases analyzed. 
We then created a generic neural model to investigate adaptive learning rules that create and maintain lognormal distributions.
We conclusively demonstrate that not only weights, but also intrinsic gains, need to have strong Hebbian learning in order to produce and maintain the experimentally attested distributions. This provides a solution to the long-standing question about the type of plasticity exhibited by intrinsic excitability.
\end{abstract}
%\section*{Keywords}
%neural coding, Hebbian learning, intrinsic excitability, rate coding, spike frequency, neural circuits, neural networks, lognormal distributions. 

\section{Introduction}

Individual neurons have very different, but mostly stable, mean spike rates under a variety of conditions \cite{SfN2006,HromadkaTetal2008}. To report on behavioral results, spike counts are often normalized with respect to the mean for each neuron. But this obscures an important question: Why do neurons within a tissue operate at radically different levels of output frequency? In order to answer this question our approach is twofold: (a) we try to document this phenomenon for different neural tissues and behavioral conditions. We also examine neural properties for their distribution, namely intrinsic gains and synaptic weights; (b) we build a very generic neural model to explore the conditions for generating and maintaining these distributions. First, we give examples for the  distribution of mean spike rates for principal neurons under spontaneous conditions, as well as in response to stimuli. We furthermore document distributions for intrinsic excitability \cite{HopfWetal2005,Mahon2012,Gunay2008} for cortical and striatal neurons, as well as synaptic weight distributions \cite{Song2005,Mason1991,Feldmeyer2006,Holmgren2003,Sayer1990,Brunel2004}.

With the current data, we show that the distribution of spike rates within any neural tissue follows a power-law distribution, i.e. a distribution with a `heavy tail'. There is also a small number of very low-frequency neurons, so that we have a lognormal distribution \cite{HromadkaTetal2008}. This lognormal distribution is present in spontaneous spike rates as well as under behavioral stimulation. For each neuron, the deviation from the mean rate attributable to a stimulus is small (CV\,=\,0.3--1, standard deviation\,=\,1--4\,spikes/s), when compared to the variability in mean spike rate over the whole population (5-7-fold), cf.\ Tables~\ref{tab:1} and\ref{tab:3}.

This work refers back to data initially reported in \cite{SfN2006}. At the time, we only had data on spike rates of cortical neurons available, plus independent evidence on intrinsic properties of striatal neurons. The observation on cortical data was taken up by \cite{Shafi2007,HromadkaTetal2008}, and led to a number of papers \cite{Roxin2011,Buszaki} focusing on the power-law distribution of spike rates as a cortical phenomenon, seeking explanations in the recurrent excitatory connectivity of cortical tissue \cite{HromadkaTetal2008,Koulakov2009,Roxin2011,Wohrer2013}. However,we find the same spike rate distributions for midbrain nuclei, medium striatal neurons and cerebellar Purkinje cells, which do not have this kind of connectivity. It has even been found in the spinal motor networks of turtles \cite{Petersen2016}. We then extended the data search for intrinsic excitability and found that lognormal distributions are ubiquitous there as well, at least in cortical as well as in striatal tissues. Finally, lognormal distributions have also been found for synaptic weights \cite{Song2005,Mason1991,Feldmeyer2006,Holmgren2003,Sayer1990,Brunel2004,Ikegaya2013}. The explanation for this universal phenomenon must lie elsewhere.

For this purpose we constructed a generic model for neuronal populations with adaptable weights and gains. We initialized both weights and gains with uniform, Gaussian or lognormal distributions.We then employed either Hebbian or homeostatic adaptation rules on both.
Under a variety of conditions we could show that lognormal distributions develop from any initial distribution only with Hebbian (positive) adaptivity. Additional homeostatic adaptation stabilized learning but erased the lognormal distributions if it was stronger than Hebbian adaptation. We could even show that the widths of the distributions from the model match with the experimental data for rates, weights, and gains (Tables~\ref{tab:1} and~\ref{tab:3}) that we have available. 
Lognormal distributions can only be maintained by positive, Hebbian-type learning rules \cite{Koulakov2009}, while homeostatic plasticity alone destroys lognormal distributions \cite{Scheler2014}.

There are a number of different learning rules and variants which all follow the 'Hebbian' principle: strong activation leads to strengthening, weak activation leads to weakening \cite{Gilson2015}. STDP rules are a variant of Hebbian learning for spiking neurons, which emphasize temporal sequence, but have the same positive learning effect \cite{Song2000, Gerstner2002,Caporale2008}. It has been noticed that positive learning rules lead to run-away activation and unstable network behavior, and that they need to be counteracted by homeostatic processes \cite{Zenke2017,Tetzlaff2012} . 
We present a generalized model of synaptic learning which consists of both Hebbian (positive) and homeostatic (negative) adaptation rules  \cite{Zenke2017}, and show that positive (Hebbian) learning is necessary to establish a lognormal synaptic weight distribution. 

For intrinsic learning it has often been assumed that it may implement purely homeostatic adaptation \cite{Desai2003,Naude2013,Cannon2017,Kazantsev2012,Cudmore2008}, but see also \cite{Tully2014}. Experimental results are often inconsistent \cite{Sehgal2013,Whitaker2017,Greenhill2015,Mahon2012,PazJetal2009,CampanacEetal2013,CampanacEetal2008,FrickAetal2004}. We will present results for a lognormal gain distribution in a number of tissues. It will be shown by simulation that the same principle holds: only a Hebbian, positive learning rule is capable of maintaining lognormal distributions, while homeostatic adaptation serves to establish stability.

This finally answers the question that experimental researchers have investigated for some time: Is intrinsic plasticity mostly homeostatic, i.e. adjusts values inversely to use, or is there Hebbian, positive learning involved: when a neuron fires, does its gain increase?  The answer is that the attested distribution of intrinsic gain can only derive from a Hebbian style adjustment rule, even though additional homeostatic adaptation is possible. Intrinsic plasticity is Hebbian.

\section{Methods}
In this section we report on data collection for spike rates, intrinsic properties and synaptic weights. Secondly, we explain the simulation model we constructed to explore the generation and persistence of the attested distributions.

\subsection{Experimental data}
We analyze five data sets for spike rates from principal neurons under behavioral activation:
\begin{enumerate} 
  \item inferior temporal (IT) cortex from monkeys \cite{Hung2005a}
  \item primary auditory cortex (A1) from monkeys  \cite{Wang2005}
  \item primary auditory cortex (A1) of rats \cite{HromadkaTetal2008} 
  \item Purkinje cells in cerebellum \cite{Raman1999,Hausser1997,DeSolages2008,Roitman2005} 
  \item midbrain principal cells from inferior colliculus (IC) from the guinea pig \cite{Zohar2013}
  \end{enumerate}

\begin{figure}[h!]
\centering
\showfig{\includegraphics[width=0.59\textwidth]{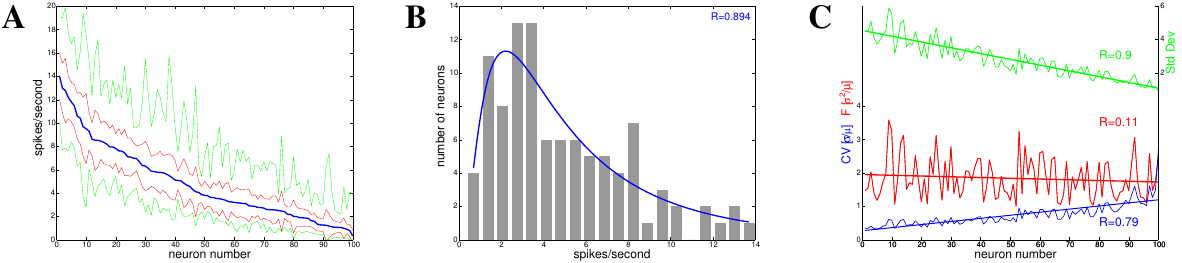}}
\caption{\textbf{Spike rate data for neurons from inferotemporal cortex (IT) in monkeys \cite{Hung2005a}.} 100 neurons, passive viewing of 77 stimuli, 10 trials (770 data points per neuron), data collected over 200ms. The data are shown for each neuron, where neurons are sorted by mean spike count.  {\bf A}: Mean spike rates (blue), standard deviation (red), and minimum/maximum absolute values (green).  {\bf B}: Mean spike rates histogram shows a lognormal distribution ($\sigma$*=2.32). 
{\bf C}: Distributions of standard deviation (green) and CV (blue) have linear slopes, with small variation. The Fano factor (red), measuring the dispersion for each neuron, is nearly constant at about 2.}
\label{IT-neurons}
\end{figure}

In monkey IT, single unit activity was recorded over 200ms for passive viewing of 77 different natural stimuli for 100 neurons, each stimulus shown 10 times \cite{Hung2005a}. This yielded 770 spike rate response data points per neuron. For these data, we show mean spike rate, standard deviation, max-min values, coefficient of variation (CV) and Fano factor (FF) (Figure~\ref{IT-neurons}), (cf. \cite{SfN2006}). What is remarkable is that the dispersion for each neuron (variance related to mean, FF) is fairly constant, and not related to the rank of a neuron as high or low-frequency. In other words, neurons have roughly similar behavioral responses relative to their average spike rate. For this reason,many behavioral experiments have reported percentage of increase/decrease of spiking as the relevant parameter.

\begin{figure}[h!]
\begin{center}
\showfig{\includegraphics[width=0.99\textwidth]{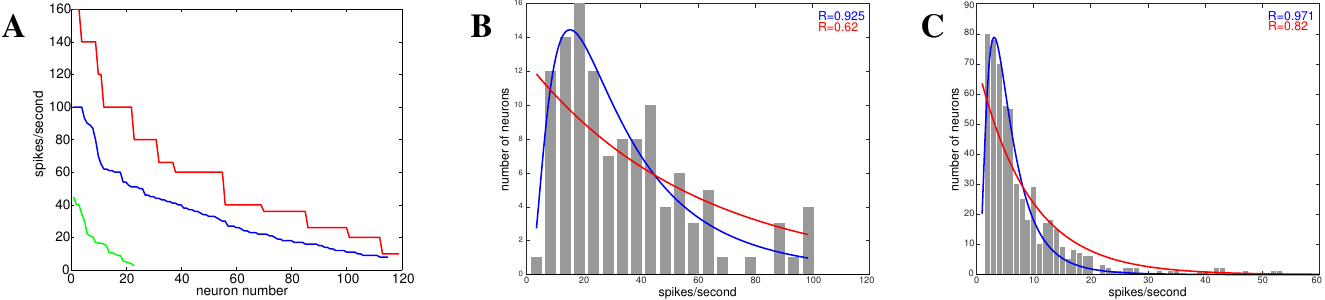}}
\end{center}
\caption{
\textbf{Responses to pure tones in primary auditory cortex from awake monkeys \cite{Wang2005} and firing rates from primary auditory cortex in rats \cite{HromadkaTetal2008}.}
{\bf A}: Distribution of spike rates to a 50ms tone ($n=119$, red)
100ms tone ($n=115$, blue), or 200ms tone ($n=23$, green) 
in primary auditory cortex in monkeys \cite{Wang2005}.
{\bf B}: Histogram for spike rate distribution for the 100ms tone
response ($n=115$)
fitted by an exponential (red) or lognormal (blue) distribution \cite{Wang2005}.
{\bf C}: Spontaneous spike rate distribution from primary auditory cortex in unanesthetized rats \cite{HromadkaTetal2008} 
fitted by an exponential (red) or lognormal (blue) distribution.
Note that the spontaneous firing rates are much lower and narrower distributed than evoked spikes in response to stimuli at short time scales (B), but that they still follow a lognormal distribution.}
\label{A1-neurons}
\end{figure}

Additionally, we show data from primary auditory cortex (A1) from awake monkeys, which were recorded for spike responses to a 50ms, 100ms, or 200ms pure tone (\cite{Wang2005}, Figure~\ref{A1-neurons}A and B) 
and data for spike rates from the primary auditory cortex of rats for four different conditions, which were recorded as cell-attached \textit{in vivo} recordings (\cite{HromadkaTetal2008}, Figure~\ref{A1-neurons}C).  

\begin{figure}[h!] 
  \begin{center}
\showfig{\includegraphics[width=0.99\textwidth]{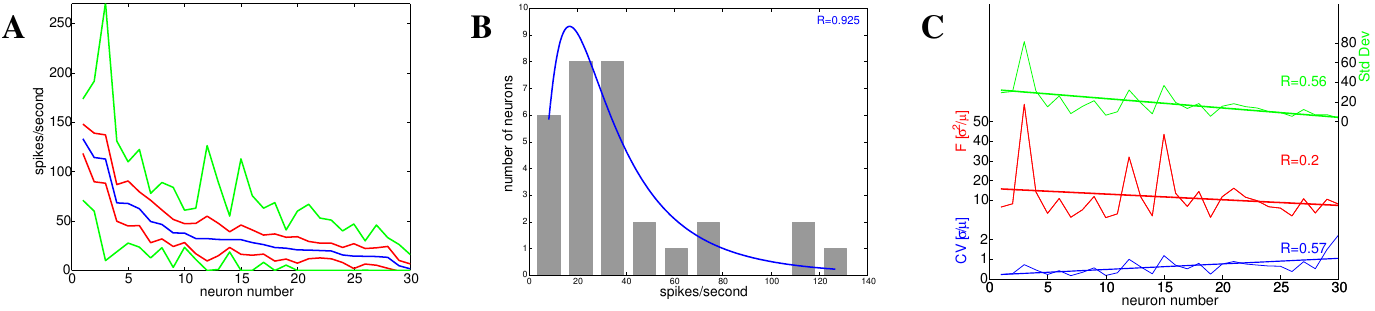}}
 \end{center}
 \caption{
\textbf{Neuronal response to binaural stimulation for inferior colliculus of the guinea pig  \cite{Zohar2013} ($n=30$), data collected over 100 ms, 200-500 trials. }
The data are shown for each neuron, with neurons sorted by mean spike count.
{\bf A}: Mean spike rates (blue), standard deviation (red), and 
minimum/maximum absolute values (green). 
{\bf B}: Mean spike rates histogram shows a lognormal distribution. 
 {\bf C}: Distributions of standard deviation (green), CV (blue) and FF (red).  Again, the dispersion is fairly constant.}
  \label{zohar}
\end{figure}

For midbrain nuclei neurons (IC), we re-analyzed spike rates in response to tones (for 200ms after stimulus onset) under variations of binaural correlation \cite{Zohar2013}. The frequency ranking of neurons by mean spike rate, standard deviation, min-max values, CV and FF are shown in (Figure~\ref{zohar}A-C). CV and FF are similar to the cortical data. 

\begin{figure}[h!] 
  \begin{center}
  \showfig{\includegraphics[width=0.7\textwidth]{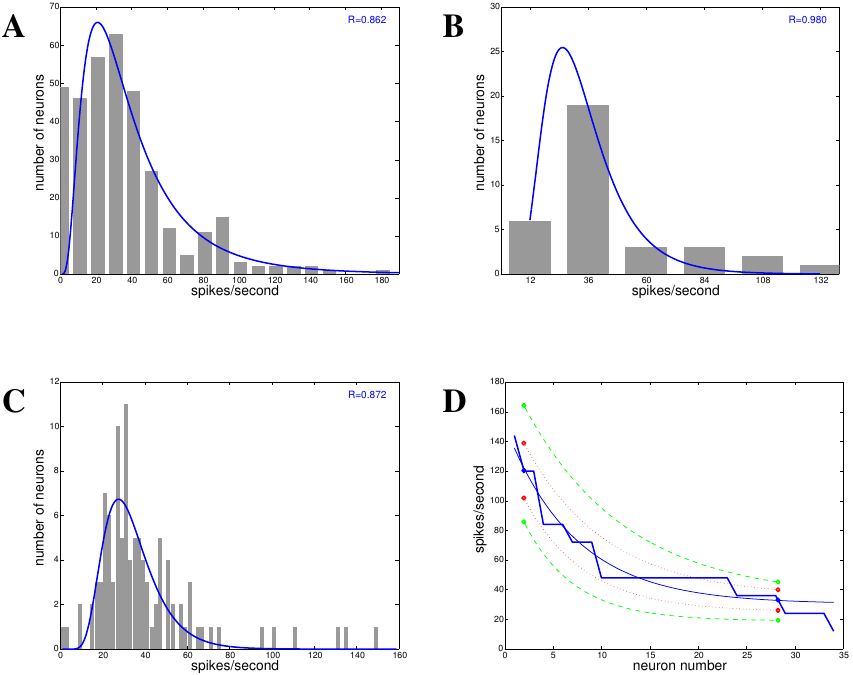}}
  \end{center}
 \caption{
\textbf{Spike rates for cerebellar Purkinje cells from rats    \cite{Raman1999,Hausser1997,DeSolages2008,Roitman2005}.} 
{\bf A:} Data for single spikes for Purkinje cells recorded from anesthetized rats (spontaneous \textit{in vivo}) \cite{DeSolages2008}($n=346$). 
{\bf B}: Data for spike frequencies of isolated cell bodies of mouse Purkinje cells \textit{in vitro}
  \cite{Raman1999} ($n=34$). 
{\bf C}: Spontaneous spike rates for Purkinje cells in slices ($n=106$) \cite{Hausser1997}. 
{\bf D}: Spike counts per neuron from  \cite{Raman1999} ($n=34$), together with variability data 
from \cite{Roitman2005} ($n=2$).
}
 \label{cerebellar}
\end{figure}

Data from GABAergic cerebellar Purkinje cells offer some difficulty for this analysis since they have regular single spikes at high frequencies, and in addition, calcium-based complex spikes \cite{DeSolages2008}. Complex spike rates however are low ($<1Hz$). This can therefore be regarded as a form of multiplexing, with two separate codes, where single spike rates can be separately assessed in their distribution. Here we report data for single spikes from \textit{in vivo} recordings in anesthetized rats \cite{DeSolages2008} (Figure~\ref{cerebellar}A)and data from spontaneous spiking (in the absence of synaptic stimulation) under \textit{in vitro} conditions (\cite{Raman1999, Hausser1997}, Figure~\ref{cerebellar}B and C). 

In order to show values for standard deviation and variance, data for two Purkinje cells from a behavioral experiment, \cite{Roitman2005}, i.e. single spike rates during arm movements from monkeys, have been added to the ranking of spontaneously firing neurons by mean spike rate from \cite{Raman1999} (Figure~\ref{cerebellar}D).

The logarithmic (heavy-tailed) distribution of spike rates is evident under all conditions.

\begin{figure}[h!]
\begin{center}
\showfig{\includegraphics[width=0.99\textwidth]{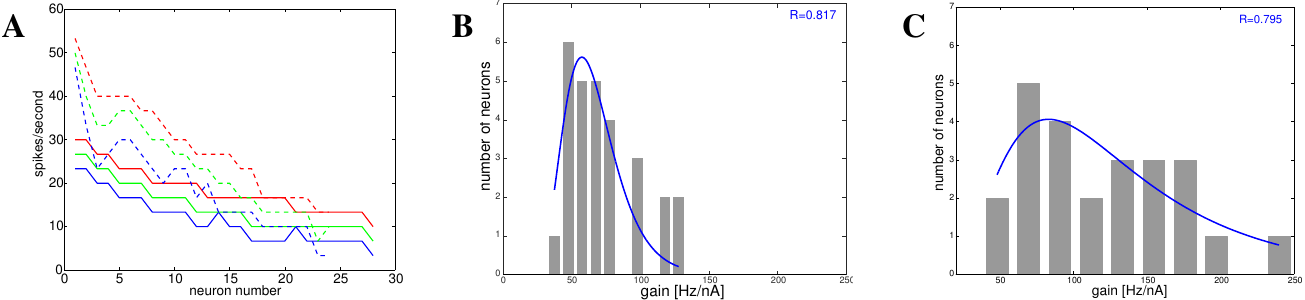}}
\end{center}
\caption{\textbf{Spike rate and gain distributions in basal ganglia \cite{HopfWetal2005}.}
{\bf A}: Spike rate in response to a 300ms constant current pulse at 
 180pA (blue), 200pA (green), and 220pA (red) for neurons in dorsal striatum ($n=28$, solid lines) and 
nucleus accumbens shell ($n=24$, dashed lines). 
{\bf B}: Gain (Hz/nA) for neurons in dorsal striatum ($n=28$).
{\bf C}: Gain (Hz/nA) for neurons in NAcb shell ($n=24$).
}
\label{gain-neurons0}
\end{figure}

\begin{figure}[h!]
\centering
\showfig{\includegraphics[width=0.99\textwidth]{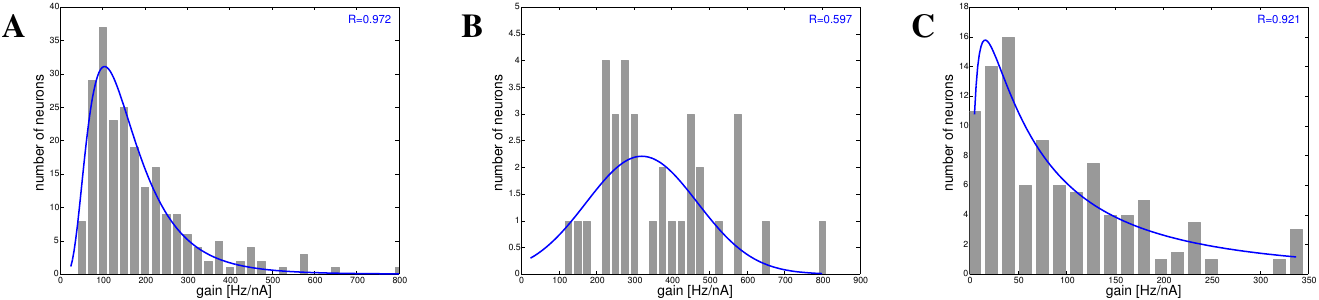}}
\caption{\textbf{Gain [Hz/nA] for cortical and striatal neurons.}
{\bf A}: Gain for all types of cortical neurons \textit{in vivo} ($n=220$) \cite{Nowak2003}.
{\bf B}: Gain for fast spiking cortical neurons only ($n=33$) \cite{Nowak2003}.
{\bf C}: Gain for neurons in globus pallidus (GP) in response to a +100pA current 
pulse ($n=145$) \cite{Gunay2008}.
}
\label{gain-neurons}
\end{figure}

The distribution of spike rates for neurons spiking in the absence of synaptic input shows that there are differences in the intrinsic excitability of neurons. To further explore this we looked at three additional datasets, which report the action potential firing of a cell due to injected current, such as by a constant pulse. This can be defined as the neuronal gain parameter (spike rate divided by current, [Hz/nA]). 
\begin{enumerate}
  \item medium striatal neurons in slices from rat dorsal striatum and nucleus accumbens shell (NAcb shell) \cite{HopfWetal2005}, cf. \cite{SfN2006}, Figure~\ref{gain-neurons0}.
  \item cortical neurons in cat area 17 in vivo \cite{Nowak2003}, Figure~\ref{gain-neurons}A and B
  \item striatal neurons from globus pallidus (GP) from awake rats \cite{Gunay2008}, Figure~\ref{gain-neurons}C 
\end{enumerate}
 
In \cite{SfN2006}, we already presented the data from striatum, which show that the spike response to a constant current follows a heavy-tailed distribution \cite{HopfWetal2005}. Figure~\ref{gain-neurons0}A shows the spike rate in response to current pulses of different magnitude in two different areas, nucleus accumbens (NAcb) shell and dorsal striatum. Figure~\ref{gain-neurons0}B and C show the distribution of rheobase (current-to-threshold) for dorsal striatal and NAcb shell neurons. Distributions appear mostly lognormal, with the exception of the 200pA current pulse response and the data in Figure~\ref{gain-neurons0}C, which appear normally distributed.

We extend this dataset by recordings  
from different types of cortical neurons in cat area 17 \textit{in vivo} (\cite{Nowak2003}, Figure~\ref{gain-neurons}A and B) and from GP in awake rats (\cite{Gunay2008}, Figure~\ref{gain-neurons}C).

A lognormal distribution of intrinsic gain is clearly apparent, except for fast-spiking interneurons, which, however, may be a sampling error (n=33).

\begin{figure}[h!] 
  \begin{center}
\showfig{\includegraphics[width=0.79\textwidth]{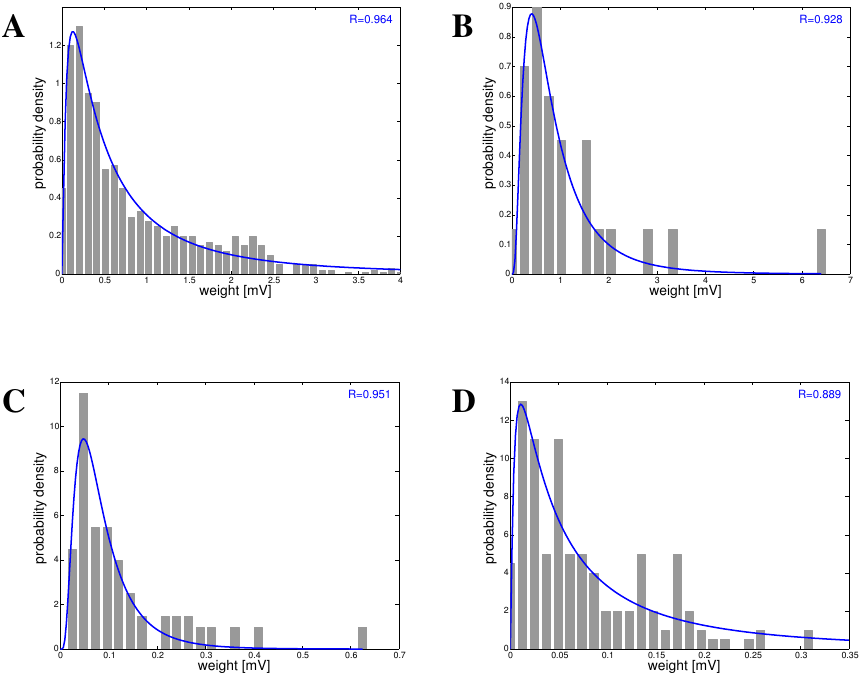}}
  \end{center}
  \caption{
\textbf{Strengths of EPSPs in cortex, hippocampus and cerebellum \cite{Sjostrom2001,Song2005,Frick2008,Sayer1990,IsopeBarbour2002}.}
	{\bf A}: Cortex: Deep-layer (L5) pyramidal-pyramidal cell connections \cite{Sjostrom2001,Song2005}.
	{\bf B}: Cortex: Deep-layer (L5) pyramidal-pyramidal cell connections \cite{Frick2008}.
	{\bf C}: Hippocampus: CA1 to CA3 connections \cite{Sayer1990}.
	{\bf D}: Cerebellum: Granule cells to Purkinje cells \cite{IsopeBarbour2002}.
}
  \label{weights}
\end{figure}

\begin{figure}[h!]
\begin{center}
\showfig{\includegraphics[width=0.8\textwidth]{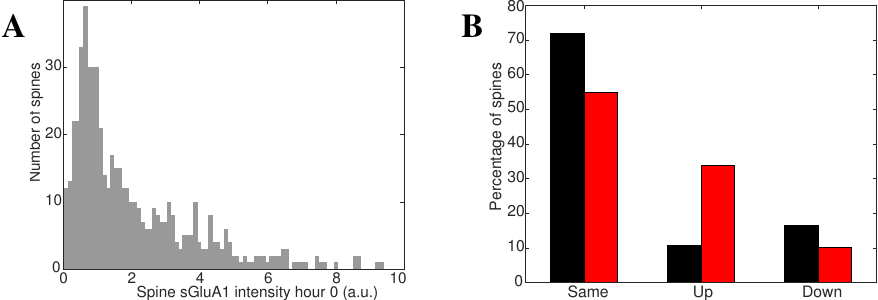}}
\end{center}
\caption{
\textbf{AMPA subunit distribution as a marker of synaptic weight.} {\bf A:} Expression of labeled GluA1 AMPA receptor subunit in layer 2/3 mouse barrel cortex in vivo follows a lognormal distribution ($\sigma^*= 2.59, \mu^* = 0.32$, $n=560$). {\bf B:} GluA1 density for control (black) and after 1 hour whisker stimulation (red). Stimulation leads to an increase of GluA1 in $\approx$~30\% of neurons   \cite{Huganir2015}.}
\label{fig:huganir}
\end{figure}

Synaptic weight distributions have been investigated starting with
 \cite{Sayer1990} in hippocampus by measuring EPSC magnitude \cite{Sjostrom2001,Song2005,VanRossum2000,Frick2008,IsopeBarbour2002} (Figure~\ref{weights}). There is also a review paper available \cite{Barbour2007} to summarize the findings. Recently, the expression of AMPA receptor subunit GluA1, which is correlated with spine size, has also been measured \cite{Huganir2015}, Figure~\ref{fig:huganir}. We used five datasets from cortex, hippocampus and cerebellum:
\begin{enumerate}
\item EPSPs for deep-layer pyramidal-pyramidal cell connections in rat visual cortex \cite{Sjostrom2001,Song2005}
\item EPSP amplitudes for deep-layer  excitatory neuron connections in somatosensory cortical slices of juvenile rats \cite{Frick2008}
	\item EPSP amplitudes for CA1 to CA3 connections in guinea pig hippocampal slices \cite{Sayer1990}
	\item EPSCs for granule cells to Purkinje cells in  adult rat cerebellar slices \cite{IsopeBarbour2002}
\item labeled GluA1 AMPA receptor subunit in mouse somatosensory barrel cortex \cite{Huganir2015}
\end{enumerate}
 
In \cite{Song2005}, EPSP magnitude was measured for L5 pyramidal neurons in slices from rat visual cortex, averaged over 45-60 responses, and peak amplitude recorded, (Figure~\ref{weights}A). 
Similar data were used in \cite{Frick2008} for slices from a single barrel column in rats (Figure~\ref{weights}B). A 30-fold variation of coupling strength was noted. In \cite{Sayer1990}, EPSPs between CA3 and CA1 in hippocampal slices were recorded, by detecting somatic membrane potential changes in response to presynaptic neuron stimulation (Figure~\ref{weights}C). There are also synaptic weight data on granule cell to Purkinje cell connections \cite{IsopeBarbour2002,Brunel2004,Barbour2007}, which show a similar distribution, but have an order of magnitude weaker connections than cortical connections (Figure~\ref{weights}D). Finally, a different type of evidence was obtained in \cite{Huganir2015}, namely labeling for a subunit of AMPA receptors in layer 2/3 mouse barrel cortex\textit{ in vivo} both before and after whisker stimulation. The AMPA intensity is distributed lognormally over the spines, corresponding to the observations on the strengths of EPSPs. It is noticeable that stimulation leads to increase of on average 200\% (two-fold) in about 30\% of spines \cite{Huganir2015}. Yet as we know, over time the overall distribution of synaptic strengths remains stable.
% * <molly.cranston@f1000.com> 2017-07-12T12:47:50.051Z:
% 
% Please provide a title for Figure 8.  -- DONE
% 
% ^.
For synaptic weights, just as for intrinsic gains and spike rates, lognormal distributions have been found for both EPSPs and AMPA receptor distribution in a highly consistent manner.

In many cases, the data were only available in the form of histograms. The parameters of the lognormal distribution were then obtained by fitting the data using a Nelder-Mead optimization method. A number of parameters were derived from these fits and reproduced in Tables~\ref{tab:1}--\ref{tab:3}, cf.~\ref{tables}.

\subsection{Simulation model}
\label{sec:2.2}
\begin{figure}[h!]
\begin{center}
\showfig{\includegraphics[width=0.7\textwidth]{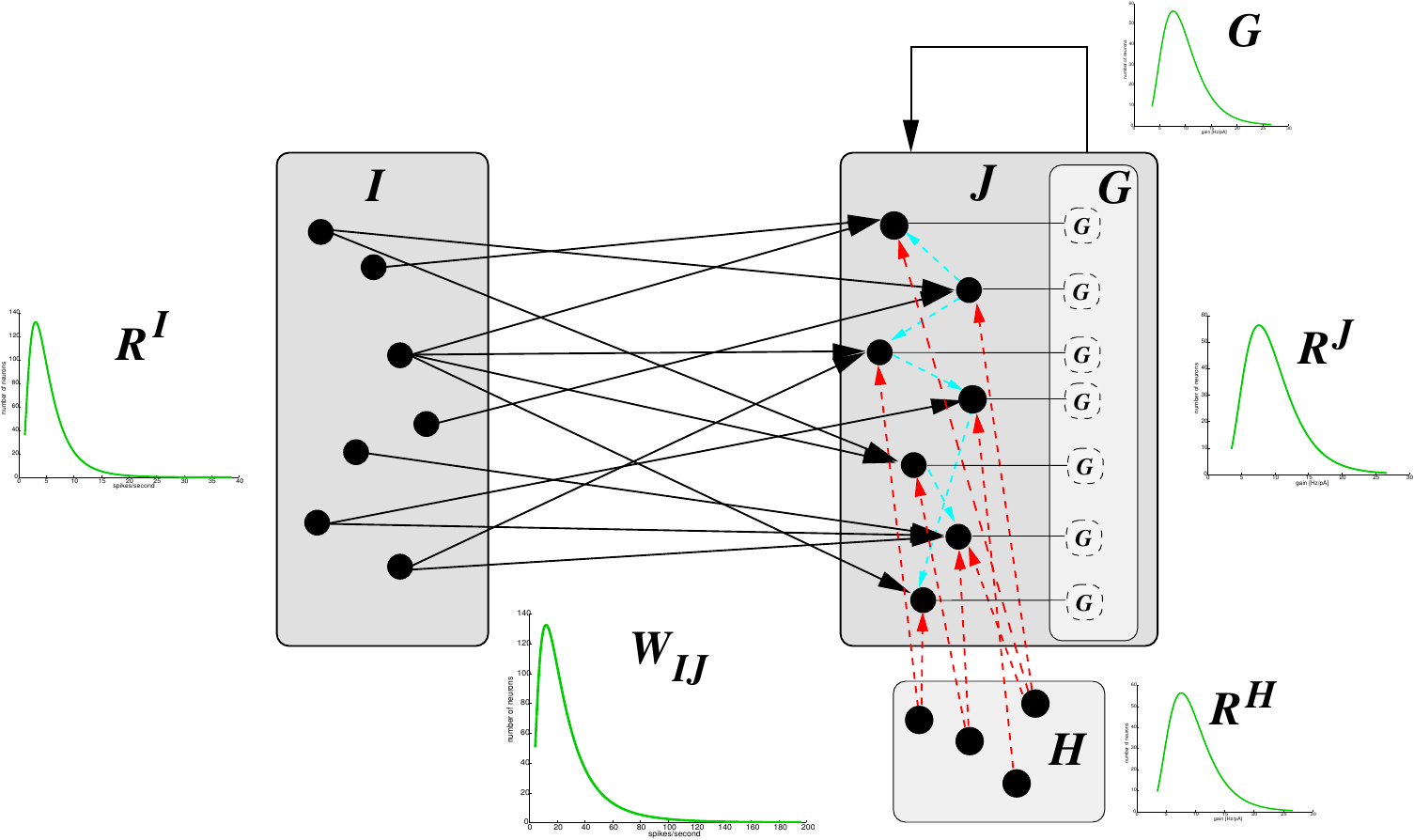}}
\end{center}
\caption{\textbf{Generic neural network model with neuron populations $I$, $J$ (excitatory, blue arrows) and $H$ (inhibitory, red arrows).} Lognormal distributions occur for gain $G$, rates $R^I$, $R^J$, $R^H$, and weight distributions $W_{IJ}$. J may have recurrent connectivity.}
\label{fig:hist-mult}
\end{figure}

Given are two neuron populations $I$, $J$ each with $n = 1000$ neurons and variable random connectivity $C$ between $I$ and $J$. $C$ determines the density between $I$ and $J$. The input population $I$ always has excitatory output onto $J$. Inhibitory input to $J$ is modeled by a population $H$ with $n=200$. The output neuron population $J$ may also have recurrent excitatory connectivity. Figure~\ref{fig:hist-mult} shows the architecture for the generic neural network used. The model (GNN) was programmed in Matlab, and is available in the public repository github (https://github.com/gscheler/GNN, DOI: https://doi.org/10.5281/zenodo.829949).

The input population $I$ is modeled according to \cite{HromadkaTetal2008} for pyramidal cortical neurons with a spike rate distribution of $\mu^* = 4.95$ and $\sigma^* = 1.98$ (Table~\ref{tab:1}). The goal is to generate a spike rate distribution $R^J$ for $J$, given a gain distribution $G$ for the target neurons, and the weight distribution $W_{IJ}$ such that $R^J$ is similar to $R^I$. 

For each neuron $j$ in $J$ the spike rate $r_j$ is calculated by applying its gain $g_j$ to the weighted sum of its connected excitatory inputs and its inhibition. $C_j$ is the set of neurons from $I$ that have excitatory connections to neuron $j$.
\begin{equation}
r_j = g_j ( \sum_{i \in C_j} w_{ij} r_i - r^H_j)
\end{equation}
where the rate $r_i$ is taken from the distribution $R^I$, $r^H_j$ from $R^H$, $w_{ij}$ from $W_{IJ}$, and $g_j$ from $G$.
$g_j$ is modeled as a factor for a linear gain function. It is possible to use a sigmoidal gain function instead, but this makes no difference for the conclusions from the model (Section~\ref{sec:generating}). The output $R^J$ may be used as input to $I$
% * <molly.cranston@f1000.com> 2017-07-12T10:32:52.450Z:
% 
% > (s.3.3)
% Is this section 3.3?  --DONE
% 
% ^.
with a matrix $W_{I,J}$ for tests of the adaptation rules.

For the adaptation of weights $W_{IJ}$ and gains $G$, we use Hebbian or homeostatic rules, as described in Section~\ref{sec:generating}. 

The system described in this way is sufficient for all the calculations on the shape of distributions used in this paper.  

\section{Results}
\subsection{Universality of lognormal distributions}
We have documented the distribution of spike rates, gains, and weights for different types of neurons (Figures~\ref{IT-neurons}-~\ref{fig:huganir}).

The distribution in all cases follows a lognormal shape. In some cases, we had data on the variability of spike rates and analyzed them for dispersion (CV, FF) under behavioral stimulation. While the fold-change from low spiking neurons to high spiking neurons is high, 5- to 7-fold, the variability for each neuron is comparatively low. It also seems to be adequately described by a percentage change over the whole population. This means that a low spiking neuron never reaches the same rate as a high spiking neuron, even when fully activated.

The similarities across neural systems are striking. For instance, in a midbrain nucleus (inferior colliculus) which is essentially an 'output' site for auditory and somatosensory cortex, spike rates are high overall \cite{Zohar2013}, nonetheless the distribution of mean spike rate, and the variability are comparable to cortical data (Figure~\ref{zohar}). Hippocampus, cerebellum and cortex vary in degree of bursting and spike irregularity \cite{Mochizuki2016}, but the rate distribution is constant. The distribution of mean spike rate is also essentially the same under spontaneous and under behavioral conditions.

Lognormal distributions were obtained by fit to the histograms obtained from data (goodness of linear fit, mean $\approx 0.92$, s. figures 1-8). The lognormal distribution is a very simple statistical distribution \cite{Limpertetal2001}, almost as simple and as universal in the description of natural processes as a Gaussian distribution (with which it is identical for small $\sigma^*$). Even though the datasets were occasionally fairly small, and more data could be added to obtain greater precision, the conclusion seems warranted that the underlying natural process is as simple and general as the multiplication of independent variables \cite{Crow88}, rather than assuming more complex processes which may lead to other exponential-family distributions. 

Lognormal rate distributions appear to be an essential property of neural tissues that occur in areas with very different neuron types and connectivity, and different absolute spike frequencies.  They are present during spontaneous activity, and under activation of a network, \textit{in vivo} as well as \textit{in vitro}. They have a counterpart in a lognormal distribution of intrinsic excitability, and lognormal synaptic connectivity. This type of distribution seems to be an essential component of the functional structure of a mature network, which is not altered by learning, plasticity, or processing of information.

\subsection{Data analysis for distributions}
\label{tables}
A lognormal distribution is characterized by parameters $\mu^*$and  $\sigma^*$. $\mu^* = e^\mu$  is the median, a scale parameter, which determines the height of the distribution. $\sigma^* = e^\sigma$ is the multiplicative standard deviation, a shape parameter which determines the width of the distribution. For distributions with small $\sigma^*$(approximately $\sigma^* <1.2$  or $\sigma< 0.182$) a lognormal distribution is essentially identical to a normal distribution. (The coefficient of variation  $CV \sim \sigma^* -1$, so that for $CV < 0.18$, a lognormal equals normal distribution.)  We collected data on spike rate, gain and synaptic weight distribution for a number of tissues in different experimental conditions (Tables~\ref{tab:1} -\ref{tab:3}). For the height of the spike rate distribution, there are known differences, e.g. with lower values for cortex ($\mu^* \approx 4.5$) and higher values for Purkinje cell ($\mu^* \approx 30$) and midbrain nuclei (cf.\ Table~\ref{tab:1}). In other words, spike rates differ between brain areas such as cortex and cerebellum by a factor of 10. 

In contrast to that, the width of spike rate distributions is more similar, with an average at $\sigma^* \approx 2.2$, with one outlier. The gain has a smaller $\sigma^*$, i.e. a more normal, less heavy-tailed distribution than the spike rate. Minus the outlier (3.46), the mean for $\sigma^*$ is only 1.86, considerably lower than the width of the spike rate distribution (Table~\ref{tab:2}). For weight distributions (Table~\ref{tab:3}), the width $\sigma^*$ is consistently larger, with an average of almost 3 (2.91). The synaptic strength ($\mu^*$) varies over at least one order of magnitude between cortex and cerebellum.

It turns out that $\sigma^*$ values are significantly different for rates, gains and weights, lowest for gains ($\sigma^* \approx 1.8$), higher for rates ($\sigma^* \approx 2.2$) and highest,($\sigma^* \approx 3$), for weights. The data that we have are not precise enough to draw quantitative conclusions, but no large distinctions are apparent between the tissues (Tables~\ref{tab:1} - \ref{tab:3}). We use a generic neural network to recreate lognormal distributions by adaptation rules and we will also show that distribution widths are structural properties which follow from general network properties.

%TABLE 1
%----------------------------------------------------------------------
\begin{table}[h!]
\centering
%\begin{tabular}{|l|r|r|r|r|r|r|}
\begin{tabular}{|l|rrrrrr|}
\hline
Tissue	 & Mean  &Median & Variance  & Width & Mode &  \\
      	 & $\mu$ & $\mu^*$ & $\sigma^2$ & $\sigma^*$ & $e^{\mu-\sigma^2}$ & n \\
\hline
 IT cortex~\cite{Hung2005a} &1.5& 4.50 & 0.71 & 2.32 & 2.2 & 100 \\
 A1 cortex~\cite{HromadkaTetal2008} &1.6 & 4.95 &0.47 & 1.98 & 3.1 &145\\
 A1 cortex~\cite{Wang2005} &3.3 & 27.11 & 0.69 & 2.29 &13.6 &263\\
 Purkinje \textit{in vitro}~\cite{Hausser1997} & 3.46 & 31.82 &0.14 & 1.46 &27.6 &106\\
 Purkinje \textit{in vitro}~\cite{Raman1999} & 3.44 & 31.19 & 0.198 & 1.56 &30.0 &34\\
 Purkinje \textit{in vivo}~\cite{DeSolages2008}  & 3.5 & 33.12 &0.47 & 1.98 &20.7 & 319\\
 Inferior colliculus~\cite{Zohar2013}  & 3.31 &27.39 & 0.90 & 2.58 & 11.13 & 30 \\ 
\hline
\end{tabular}
\caption{Statistics of spike rate distributions in different tissues.}
\label{tab:1}
\end{table}

% TABLE 2
%----------------------------------------------------------------------
\begin{table}[h!]
\centering
\begin{tabular}{|l|rrrrrrr|}
\hline

Tissue &Peak&Mean& Median & Variance & Width & Mode &	\\
       &[Hz/nA]&$\mu$[Hz/nA]& $\mu^*$ & $\sigma^2$ & $\sigma^*$ & $e^{\mu-\sigma^2}$ &	n\\
\hline
 Dorsal striatum~\cite{HopfWetal2005}	& 48& 4.24 & 69.41 & 0.36 & 1.82 & 48.3 &28\\
% * <molly.cranston@f1000.com> 2017-07-12T12:30:18.750Z:
% 
% >  Dorsal striatum	& 48& 4.24 & 69.41 & 0.36 & 1.82 & 48.3 &28\\
% >  NAcb shell
% Could you provide the two references where these data are contained?
% 
% ^.
 NAcb shell~\cite{HopfWetal2005}& 65 & 4.67 & 106.70 &0.49 & 2.01 & 65.4 & 24\\
 GP \textit{in vivo}~\cite{Gunay2008} & 6.6 & 3.4 & 29.96 &1.54 &3.46 &6.4 &146\\
 GP model~\cite{Gunay2008} & 37 &4.0 & 54.60 &0.40 & 1.88 &36.6 & 10000\\
 Cortical~\cite{Nowak2003} & 105 & 4.96 & 142.59 &0.31 & 1.75 & 104.6 &220\\
\hline
\end{tabular}
\caption{Statistics of intrinsic excitability (gain) in different tissues.}
\label{tab:2}
\end{table}

% TABLE3
%----------------------------------------------------------------------
\begin{table}[h!]
\centering
\begin{tabular}{|l|rrrrrr|}
\hline
 Tissue	 & Mean&Median&Variance& Width& Mode &\\
 	 & $\mu$&$\mu^*$&$\sigma^2$& $\sigma^*$& $e^{\mu-\sigma^2}$ & n\\
\hline
 Cortex L23~\cite{Mason1991} & -0.99 & 0.37 & 0.76 &2.39 & 0.17& 48   \\
 Cortex L23~\cite{Feldmeyer2006} & 0.25 & 1.28 & 1.41 &3.28 &0.31 & 35  \\
 Cortex L23~\cite{Holmgren2003} & -0.94 & 0.39 & 1.54 & 3.46& 0.08 & 61   \\
 Cortex L5~\cite{Sjostrom2001} & -0.56 & 0.57 & 1.47 & 3.36 & 0.13 & 1004  \\
 Cortex L5~\cite{Frick2008} & -0.31 &0.73 & 0.58 & 2.14 & 0.41 & 26  \\
 Hippocampus~\cite{Sayer1990} & -2.61 & 0.07& 0.43 & 1.93 & 0.05 & 71  \\
 Cerebellum GC-PC~\cite{Brunel2004} & -2.70 &0.07 & 1.82 &3.85 & 0.01 & 104 \\ 
 Cortex \textit{in vivo} \cite{Huganir2015} & -1.14 & 0.32 & 0.90 & 2.59 & 0.13 & 560 \\
\hline
\end{tabular}
\caption{Statistics of synaptic weight distributions in different tissues.}
\label{tab:3}
\end{table}

\subsection{Generating lognormal distributions with generic neural networks}
\label{sec:generating}
Since not only mean spike rates but also both components, intrinsic excitability and synaptic weights, have lognormal distribution, this raises the question of how the functional system that we observe is generated. It is obvious, if the data are accurate, that these are basic parameters of any simulation and need to be reproduced in any model to make it biologically realistic.

We set up a generic neural network model (cf. Section~\ref{sec:2.2}) to explore the mechanisms of generating and maintaining rate, weight and gain distributions. The model consists of a source neuron group $I$, a target group $J$, a population of inhibitory neurons $H$, which are connected with $J$,  and potentially recurrent excitation in the target group $J$. The spike rate distribution $R^I$ acts through a weight distribution $W$ onto a gain distribution $G$, where inhibition $H$ is subtracted, and a spike rate output distribution $R^J$ is produced (Figure~\ref{fig:hist-mult}).

In the simplest case, we look at two sets of neurons, the source and the target. The source sends excitatory connections to the target, and exhibits variable weights at outgoing synapses. The input 
that a target neuron receives is fed through a linear filter $G$ to produce an output rate $R^J$ according to Eq~(1). The distribution for $R^J$ depends on $G$ and $W$ as well as on $R^I$. The system is sufficient for calculations on the shape of distributions, as well as the effects of Hebbian and homeostatic plasticity.  

\begin{figure}[h!]
\begin{center}
\showfig{\includegraphics[width=0.7\textwidth]{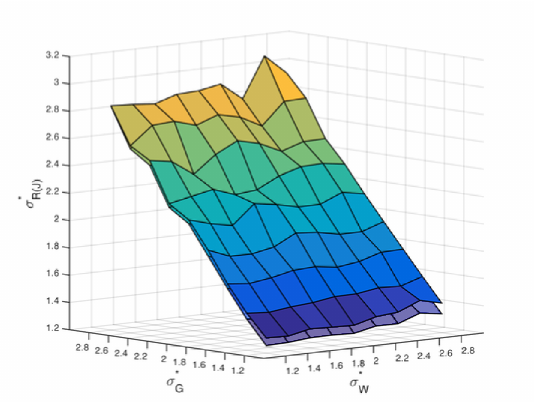}}
\end{center}
\caption{\textbf{The width of the rate distribution for $J$, $\sigma^*_{R^J}$, depends heavily on the gain $\sigma^*_G$, but not on the weight distribution $\sigma^*_{W}$.} There is a slight effect of connectivity (upper sheet C=5\%,lower sheet C=10\%).
($\mu^*_W=0.7, \mu^*_{G}=30,  
N=1000, \sigma^*_{R^I} = 2.74 , \mu^*_{R^I}=4.5$.)
}
\label{fig:res:RGW}
\end{figure}

\begin{figure}[h!]
\begin{center}
\showfig{\includegraphics[width=0.7\textwidth]{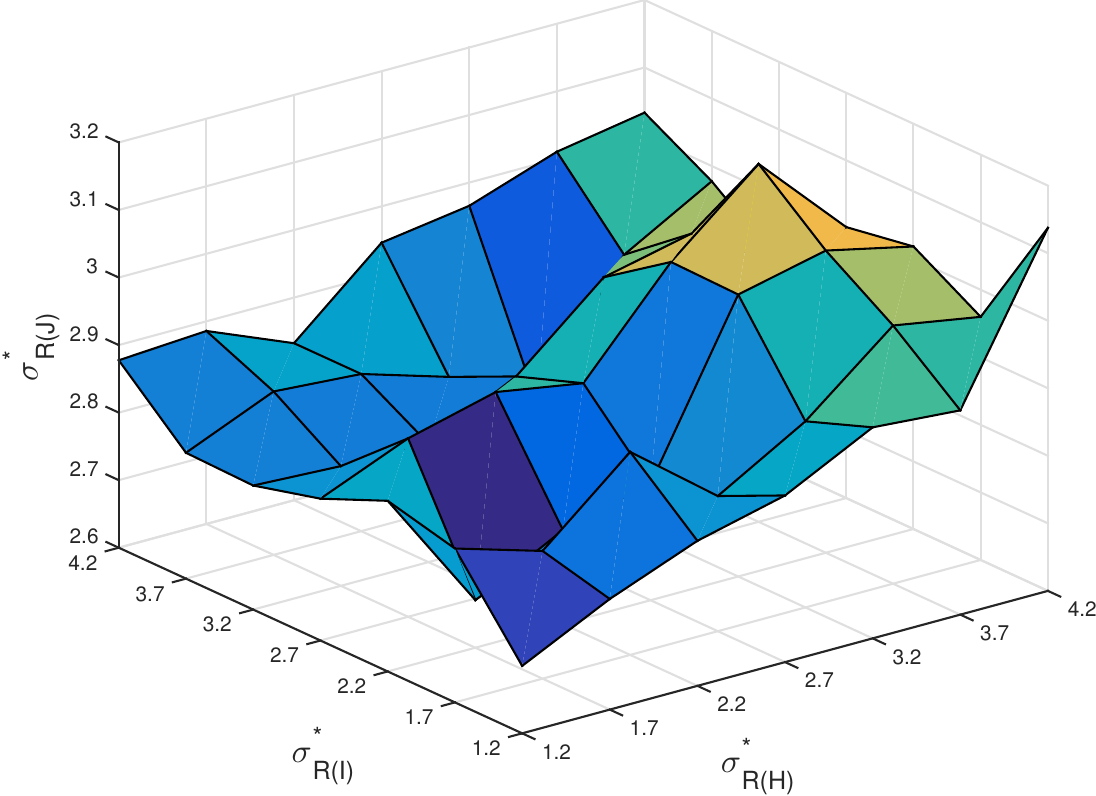}}
\end{center}
\caption{\textbf{The width of the rate distribution for $J$, $\sigma^*_{R^J}$,  does not depend on $R^I$ or $R^H$.}
($\mu^*_W=0.7$, $\mu^*_{G}=30$, $C=10\%$,  $N=1000$,  $\mu^*_{R^I}=4.5$.)
}
\label{fig:res:RHW}
\end{figure}

We have explored the dependencies between gain, weight and rate distributions in simulations.
First, we found that the width of the output spike rate distribution $R^J$ depends heavily on the gain distribution, but only slightly on the input weight distribution (Figure~\ref{fig:res:RGW}).  It does depend on the overall connectivity $C$, where $\sigma^*_{R^J}$ is wider for lower connectivity, but not very much (Figure~\ref{fig:res:RGW}). 
Secondly, the width of the output distribution $R^J$ does not depend on $R^I$ or $R^H$ either (Figure~\ref{fig:res:RHW}). The most important factor for a spike rate distribution remains the gain $\sigma^*_{G}$.
\subsection{Adaptation}
We may now ask, where do lognormal spike rate distributions come from? How is the system set up, i.e. what rules of adaptation generate lognormal distributions in weights and gains?

In the case of cortical networks, there are excitatory recurrent interactions that constitute a significant part of total input. In the case of cerebellar or striatal neurons, there are no recurrent excitatory interactions, only inhibitory interneurons and excitatory input. The generation of lognormal distributions must therefore be independent of recurrent excitation. It requires a system where continuous input shapes the weights and gains of a target network $J$.

We start with the system that we described before, with random assignment of weights and gains. We employ adaptivity for weights, and also for gains, by positive Hebbian learning, or by negative homeostatic learning. The output of $I$ is fed into $J$, and $W$ and $G$ are adaptive.  Additionally, $J$ may have excitatory recurrent connectivity, and learning takes place within the network $J$.

From any given initial spike rate distribution (Gaussian, uniform, lognormal) for $I$, we calculate $W$ assuming a positive learning (Hebbian) adjustment rule, which is dependent on input and output frequencies.  Each individual weight $w_{ij}$ is updated by 
$$ w'_{ij} = w_{ij} + \lambda w_{ij} (r^I_i\,\, r^J_j - \mu) $$
We use parameters $\lambda$ and $\mu$, such that the generated spike rate output $R^J$ is compatible in strength with the input rate $R^I$.

\begin{figure}[h!]
\begin{center}
\showfig{\includegraphics[width=0.9\textwidth]{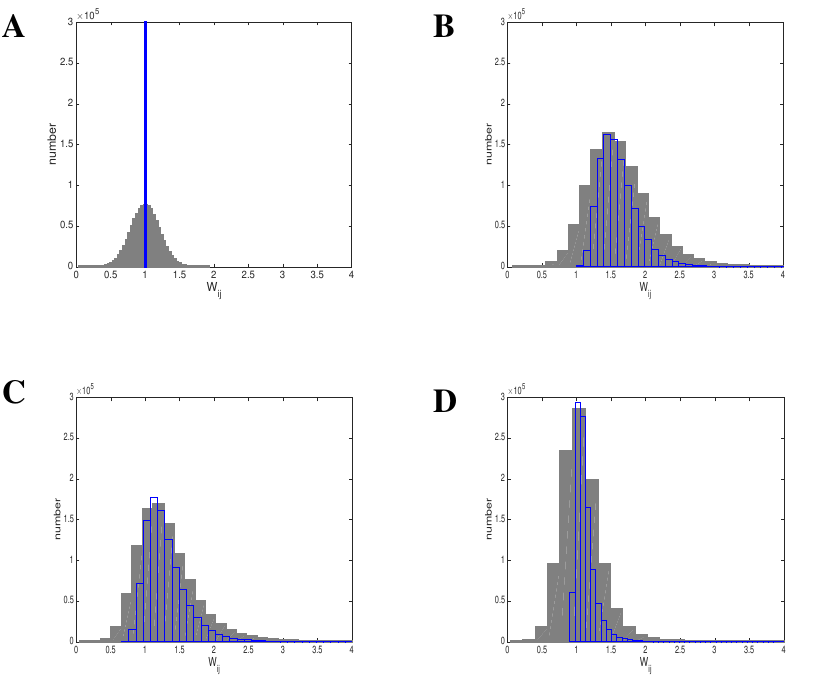}}
\end{center}
\caption{
\textbf{Hebbian learning results in lognormal weight distribution independent of gain distribution.} Given is a lognormal input rate $\sigma^*_{R^I} = 2, \mu^*_{R^I} = 4.95$.
{\bf A:} Initial weight configurations: Gaussian (grey) or uniform (blue).
{\bf B:} After Hebbian learning using Gaussian gain distribution (grey, blue as before).
{\bf C:} After Hebbian learning using uniform gain distribution (grey, blue as before).
{\bf D:} After Hebbian learning using lognormal gain distribution.
($\sigma^*_{G} = 1.37, \mu^*_{G} = 32.7$) (grey, blue as before).
}
\label{fig:res:X}
\end{figure}

Using Hebbian learning, we generate a weight distribution $W$ that is lognormally distributed, independent of the initial configuration or the distribution of the gains in the system (Figure~\ref{fig:res:X}). The lognormal distribution also develops independently of the rate distribution of the inputs,  it only develops faster with lognormal rather than normally distributed spike rate input (not shown). It makes no difference whether we use a recurrent system $J$, or a non-recurrent population $J$ with input from a population $I$ with a spike rate distribution, as long as we use a Hebbian weight adaptation rule. For the shape of the distribution, it also does not matter whether we route the output of $J$ back to $I$, or whether we use local or no recurrence.

\begin{figure}[h!]
\begin{center}
\showfig{\includegraphics[width=0.9\textwidth]{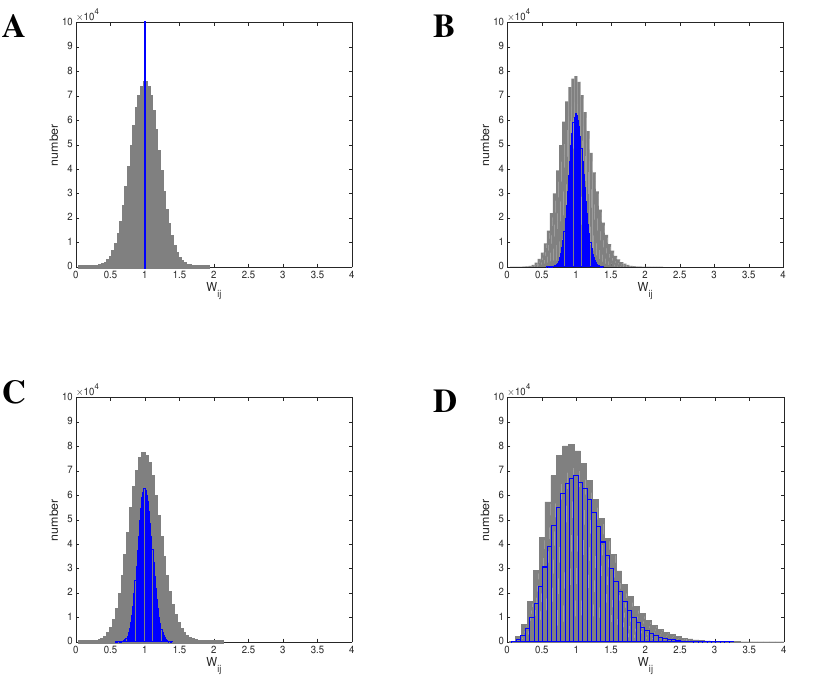}}
\end{center}
\caption{
\textbf{Homeostatic learning results in Gaussian weight distributions, independent of gain distribution.} Given is a lognormal input rate $\sigma^*_{R^I} = 2, \mu^*_{R^I} = 4.95$.
{\bf A:} Initial Configuration: Gaussian (grey) or uniform (blue).
{\bf B:} Homeostatic weight learning using Gaussian gain distribution (grey, blue as before).
{\bf C:} Homeostatic weight learning using uniform gain distribution (grey, blue as before).
{\bf D:} Homeostatic weight learning using lognormal gain distribution
($\sigma^*_{G} = 1.37, \mu^*_{G} = 32.7$) (grey, blue as before).
}
\label{fig:res:Y}
\end{figure}

To show the effect of the adaptation rule, we also used homeostatic synaptic plasticity to adjust the weights. This means that the weight is adjusted  inversely to the spike rate of input and output neurons. 
$$ w'_{ij} = w_{ij} - \lambda w_{ij} (r^I_i\,\, r^J_j - \mu) $$
In this case, it is very clear that with any input or initial configuration and any gain distribution, only a normal distribution of weights results (Figure~\ref{fig:res:Y}). Again, a lognormal input spike rate slows the process of adaptation, but the end result is the same, a normal distribution. 

Since gain distributions are also lognormal, we may ask in the same way how they develop and are maintained by plasticity rules.  We adapt the linear gain $G$ by either Hebbian or homeostatic learning. Each gain can be adjusted by a Hebbian rule
$$g_j' = g_j + \lambda g_j (r^J_j - \mu)$$ 
or a homeostatic  rule
$$g_j' = g_j - \lambda g_j (r^J_j - \mu)$$
with parameters $\lambda$ and $\mu$.

\begin{figure}[h!]
\begin{center}
\showfig{\includegraphics[width=0.99\textwidth]{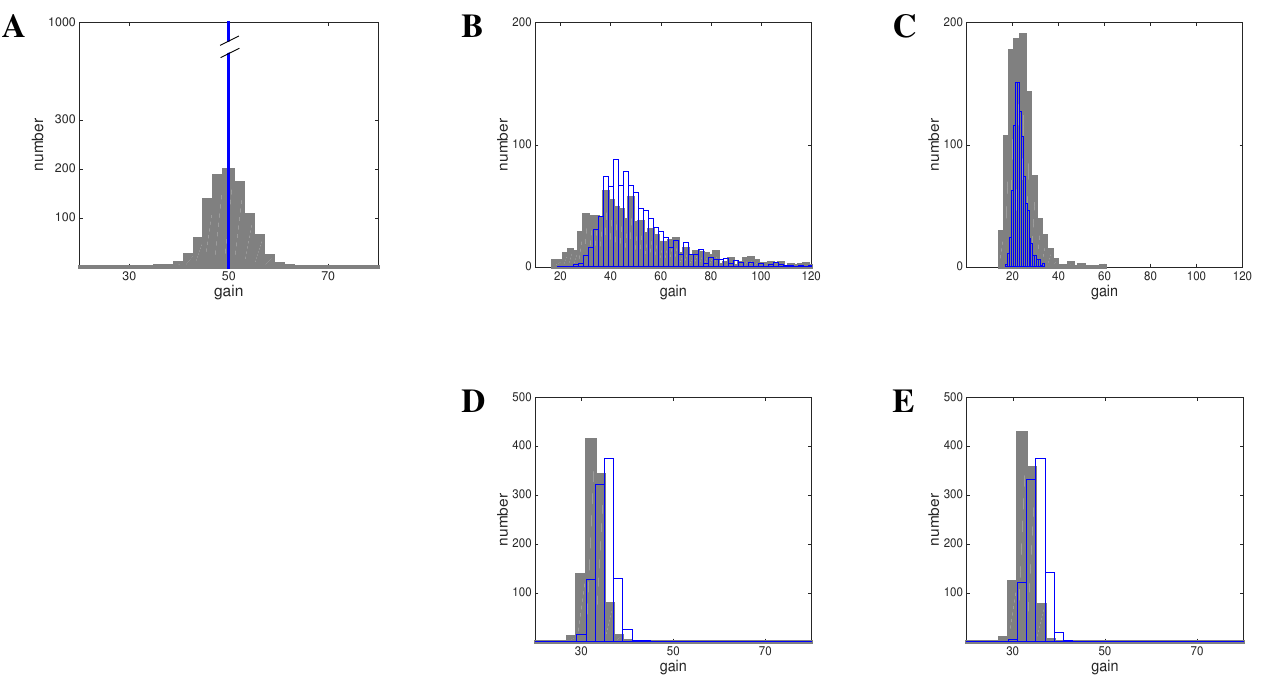}}
\end{center}
\caption{\textbf{Hebbian or homeostatic gain learning determine lognormal or Gaussian outcome.} Given is a lognormal input rate $\sigma^*_{R^I} = 2, \mu^*_{R^I} = 4.95$.
{\bf A:} Initial Configuration: Gaussian (grey) or uniform (blue).
{\bf B and C:} Hebbian learning using lognormal or Gaussian weights, resulting gain distribution is lognormal.
{\bf D and E:} Homeostatic learning using lognormal or Gaussian weights, resulting gain distribution is Gaussian.
}
\label{fig:res:Z}
\end{figure}

We start with uniform or normally distributed $G$ in an environment where $W$ is lognormal, normal or uniform, and $R^I$ is normal or lognormal. If we adapt only $G$ for any initial configuration, using any distribution for $R^I$, including the lognormal distribution, and a lognormal or normal weight distribution, we arrive at a normal distribution for $G$ with homeostatic learning and a lognormal distribution with Hebbian learning (Figure~\ref{fig:res:Z}).

\begin{figure}[h!]
\begin{center}
\showfig{\includegraphics[width=0.6\textwidth]{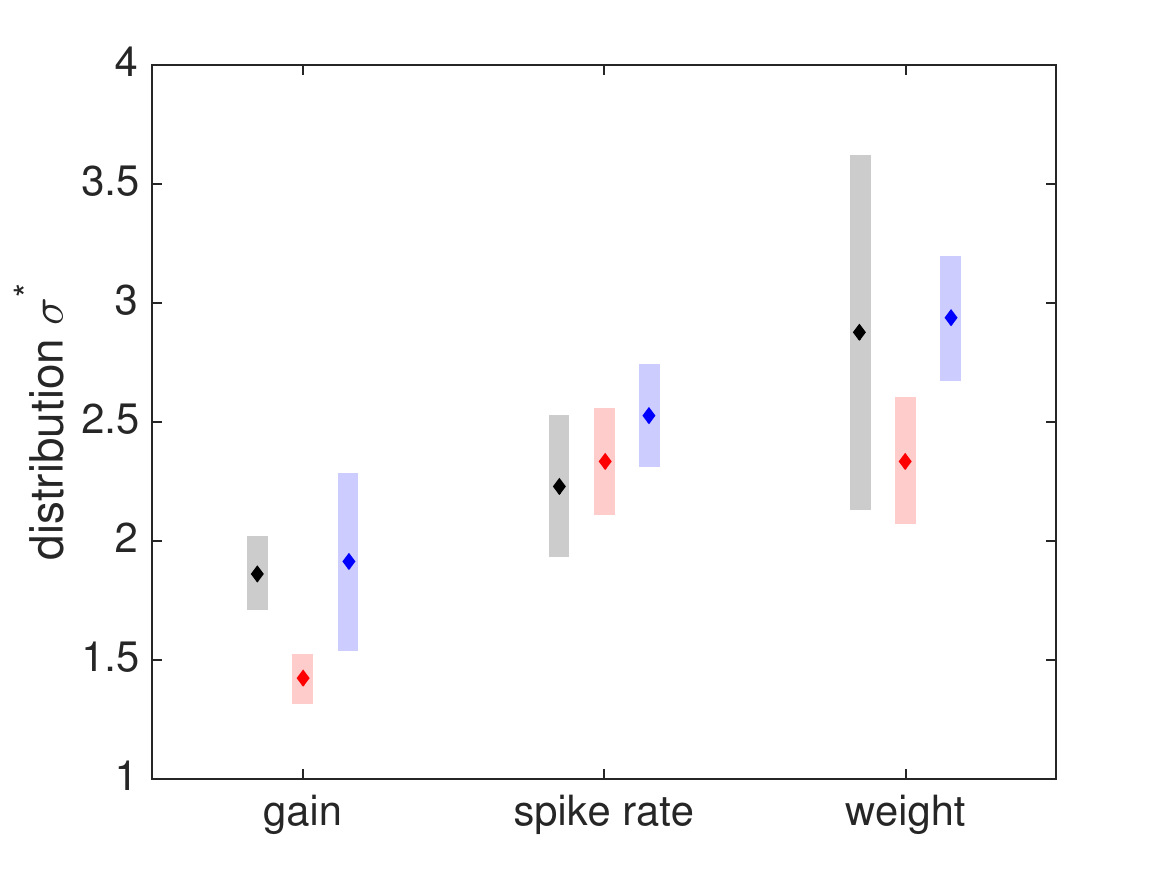}}
\end{center}
\caption{
\textbf{Experimental and generated distribution widths for spike rates, gains and weights. }Grey, experimental measurements (s. Tables); red, generated with 100\% Hebbian learning; or blue, 80\% Hebbian and 20\% homeostatic learning combined. The basic distinction in distribution width for gains, rates, weights is reproduced with Hebbian learning, additional homeostatic learning matches experimental values best.}
\label{fig:res:Z1}
\end{figure}

Lognormal distributions develop from Hebbian plasticity, and homeostatic plasticity generates only normal distributions. The explanation lies in the nature of random statistical events, which generate normal distributions when the underlying mechanisms are sums of many small events, but lognormal distributions when the underlying mechanisms are multiplicative \cite{Limpertetal2001}. 
We also wanted to understand the observed widths of the distributions. We hypothesized that the differences for $\sigma^*$ between $W$, $R$ and $G$ result from the network structure. Accordingly we started a simulation with initial uniform values for $G$ and $W$ and Hebbian update rules using the same learning rate $\lambda$ for both (Figure~\ref{fig:res:Z1}). We find that gain, rate and weight distributions match the experimental values, and that this is true for any tested constellation. We also found that Hebbian learning alone quickly escalates values, which develop exponentially, and that additional rounds of homeostatic adaptation are required to stabilize the system. Homeostatic learning pushes the system back towards a normal distribution.

Our data, in the most general way, allowing for various conditions and architectures, show that Hebbian learning is required both for intrinsic gain and for weights in order to generate the attested lognormal distributions. This is an interesting result, because it shows that we need prominent Hebbian intrinsic learning to explain the gain distributions that we find experimentally. Intrinsic learning is not just homeostatic adaptation, it follows the same rules as synaptic weight learning.  

\section{Discussion}
\subsection{Universality of lognormal distributions}
Spike rates of neurons seem to be universally distributed according to a lognormal distribution, with many neurons at low spike rates, and a small number at successively higher spike rates (heavy-tail) \cite{SfN2006}. The same distributions are found for synaptic weights \cite{Song2005}, and intrinsic properties associated with excitability (gain) \cite{SfN2006}. The neurons that we reported on are of very different types, and they are embedded in different kinds of connectivity.  Medium spiny neurons and Purkinje cells are GABAergic (inhibitory), while cortical and IC neurons are glutamatergic (excitatory), but this is not reflected in a distinct spike rate distribution. They also fire with very different average spike rates. IC neurons operate at very high frequencies, and Purkinje neurons at much higher frequencies than cortical or striatal projection neurons. But they all have the same spike rate distribution. It has been suggested \cite{HromadkaTetal2008} that lognormal spike distributions are a feature of cortical tissue and arise from strong excitatory recurrent connectivity, but this is experimentally not substantiated nor is it theoretically necessary.  While cortical pyramidal neurons exist in a heavily excitatory recurrent environment, medium spiny neurons, cerebellar Purkinje cells and IC neurons act mostly in a feed-forward way, i.e. they don't have significant recurrent excitatory (glutamatergic) connectivity. 
% * <molly.cranston@f1000.com> 2017-07-12T10:45:44.156Z:
% 
% > Hromadka
% Should this be a reference? --DONE
% 
% ^.

Beyond spike rate distribution, we also gathered data on weight and gain distributions.  Again the observation of lognormal distributions is ubiquitous. We find synaptic weight distributions for cortex \cite{Song2005} and cerebellum that are lognormal, with characteristic width of distributions. For intrinsic properties, striatal projection neurons and cortical neurons \cite{Nowak2003} show responses to constant current and current-to-threshold (gain) distributions, which again appear lognormally distributed, with smaller widths than spike rate distributions.

Our models show that lognormal distributions arise even in a purely input-output environment, and that they are a result of Hebbian learning of weights and gains, quite independent of the overall magnitude of the spike rates.

\subsection{Generating lognormal distributions}
Mean spike rates, as well as intrinsic excitability and synaptic weights, have lognormal distributions. 

It has often been assumed that variability in intrinsic excitability is a source of noise in neural computation \cite{Rudolph2003}, even though others have argued that intrinsic variability contributes to neural coding \cite{Scheler2004,StemmlerKoch1999} and that intrinsic plasticity follows certain rules \cite{Scheler2014}. An excellent overview of the experimentally attested forms of intrinsic plasticity is contained in \cite{PazJetal2009}, cf. \cite{Debanne2009,CampanacDebanne2007,DaoudalDebanne2003}. Many other detailed observations are contained in \cite{CampanacEetal2013,CampanacEetal2008,FrickAetal2004,CarvalhoTP2009}. 

Recently, Mahon and Charpier \cite{Mahon2012} have shown that intrinsic excitability is stable in individual neurons under control conditions, while stimulation protocols (e.g. in barrel cortex of anesthetized rats) change intrinsic excitability by at least 50-100\%. However, the conclusions drawn from the experimental research are often contradictory. Intrinsic plasticity is sometimes assumed to act in a negative, homeostatic way, i.e. opposite to synaptic plasticity \cite{Mahon2012}, but sometimes in a 'Hebbian', positive way, i.e. cooperative with synaptic plasticity \cite{FrickAetal2004,Mahonetal2003}. There is evidence for (short-term) negative or homeostatic plasticity, which has been previously investigated \cite{Mahon2012}.  
% * <molly.cranston@f1000.com> 2017-07-12T10:47:22.490Z:
% 
% > Mahon
% Please include the reference.
% 
% ^.

Our work has now shown that any kind of neural system with linear gains requires positive, Hebbian intrinsic plasticity to produce and maintain a lognormal distribution of gains. We also could show that the observed widths of the distributions, i.e.  the differences for $\sigma^*$ between $W$, $R$ and $G$, naturally result from the network structure and are built into the system simply by Hebbian adaptation.  

Lognormal distributions may arise as stable properties of the system during early development (the set-up of the system), i.e. before actual pattern storage or event memory develops, and they are maintained during processing by a Hebbian type of positive adaptation events. Homeostatic plasticity consists in downregulating gains or weights with increases in firing rates. Purely homeostatic learning results in normal distributions, and erases existing lognormal distributions. By combining homeostatic and Hebbian adaptation we can achieve and maintain stable lognormal distributions. 

\subsection{Why logarithmic coding schemes}
A lognormal distribution means that values are normally distributed on a logarithmic scale. From an engineering perspective, basic Hebbian plasticity for synapses and intrinsic properties is sufficient to generate stable logarithmic distributions.  If there is random variation of multiplicative events, as in Hebbian plasticity, a lognormal distribution will be the result \cite{Limpertetal2001}. 

This is related to principles of sensory coding, where logarithmic scale signal processing enhances perception of weak signals, while also being able to respond to large signals - effectively increasing the perceptual range compared to linear coding \cite{Buszaki}.  In an interconnected network logarithmic coding may turn into a property for the access of representations.  Feature clusters, or event traces could be accessed by targeted connections to the top-level neurons, which then activate lower level neurons in their immediate vicinity.  By accessing high frequency neurons preferentially, a whole feature area can be reached, and local diffusion will provide any additional computation. Similarly, the results of a local computation can be efficiently distributed by high frequency neurons to other areas. Fast point-to-point communication using only high frequency neurons may be sufficient for fast responses in many cases. Scale-free networks in general support synchronization, which is also a useful feature for rapid information transfer and access \cite{Scheler2005}.

Recently, publications \cite{Omura2015,Nigam2016} have shown that there is indeed a difference between high-frequency and low-frequency neurons in their connectivity: high-frequency neurons have short delays, strong connections, and directed targets, while low-frequency neurons have long delays, weak connections and diffuse targets.

The lognormal distribution of spike rates has significant implications for neural coding. Logarithmic spike rates are coupled with linear variance for responses to behavioral stimulation. In other words, the greatest part of the coding results already from the frequency rank of the neuron itself, such that high frequency neurons have the largest impact. A fixed mean rate for each neuron allows stable expectation values for network computations. 

Logarithmic, hierarchical coding does not need to be sparse. The low frequency neurons may matter the most in terms of input response. With lognormal synaptic weight distributions, if strong synapses are kept stable, they may transmit an input neuron's mean firing rate to targets and in this way provide stability to the system. All other synapses could be arbitrary. This would allow for continued pattern learning to be implemented by the bulk of low weight synapses, while the framework of neuronal interactions, e.g., the ensemble structure, could be unchanged.  Such a division of labor between strong synapses and weaker ones could have many advantages in a complex, modular network. 

Experimental data have often shown that sampling of neuronal responses from a large population ($10^5$ or more neurons), which become activated at 30\% or more, yields accuracy for a stimulus already for small samples (100-200 neurons or 1-2\%) (e.g., \cite{OConnor2010}). We may suggest that this happens when we sample from a highly modular structure, and we have been able to replicate the effect with lognormal networks \cite{Scheler2016}.

\section{Conclusions}
In our earlier work \cite{SfN2006}, we found that intrinsic excitability manifested by spike response to current injection and rheobase \textit{in vitro} for dorsal striatal and nucleus accumbens neurons seems to have the same distribution as the firing rate in cortex under \textit{in vivo} conditions. Approximately at the same time, \cite{Song2005} had observed a heavy-tailed distribution of synaptic weights in cortical tissue.

In this article, we have done three things: (a) collected data to show that rate, weight and gain distributions in different brain areas all follow a heavy-tailed, specifically a lognormal distribution; (b) created a generic neural network model to show that these distributions arise from Hebbian learning, and specifically that intrinsic plasticity must be Hebbian as well; and (c) shown that the width of the distributions, as experimentally attested, arise naturally from the network structure and the role of its components, in a very robust way. We have also discussed what the lognormal distribution means for neural coding: a division of labor between fast transmission by high-frequency neurons and low-level computation by low-frequency neurons in a modular structure, and possibly a division of labor between stable components (strong synapses, high-frequency neurons) and more variable components (weak synapses, low-frequency neurons). 

%\section{Software and data availability}

%\TODO{The GNN simulation software was programmed in Matlab, and is available in GitHub: https://github.com/gscheler/GNN}
%
%Archived source code as at time of publication: https://doi.org/10.5281/zenodo.82994949

%OSS approved license: Apache 2.0.

%All the data required for re-analysis of the study have been referenced throughout the manuscript.
% * <molly.cranston@f1000.com> 2017-07-12T13:06:35.366Z:
% 
% > Data availability
% Please complete this section (see email).
% 
% ^.
%The data are available as supplementary file SupplA. (dataSchelerlognormal.txt).

{%\small
\bibliographystyle{nat}

}
%===============================================================

\end{document}